\documentclass[acus]{JAC2000}

\usepackage{graphicx}

\begin{document}
\title{\bf {USE OF COHERENT TRANSITION RADIATION TO SET UP THE APS RF
THERMIONIC GUN TO PRODUCE HIGH-BRIGHTNESS BEAMS FOR SASE FEL
EXPERIMENTS} 
\thanks{Work supported by U.S. Department of Energy,
Office of Basic Energy Sciences, under Contract No. W-31-109-ENG-38.}}

\author{N. S. Sereno, M. Borland, A. H. Lumpkin, \\ Argonne/APS,
Argonne, IL, 60439-4800, USA}

\maketitle

\begin{abstract} 
We describe use of the Advanced Photon Source (APS) rf thermionic
gun~\cite{Ref0}, alpha-magnet beamline, and linac~\cite{Ref1} to
produce a stable high-brightness beam in excess of 100 amperes peak
current with normalized emittance of 10 $\pi$ mm-mrad.  To obtain peak
currents greater than 100 amperes, the rf gun system must be tuned to
produce a FWHM bunch length on the order of 350 fs.  Bunch lengths
this short are measured using coherent transition radiation (CTR)
produced when the rf gun beam, accelerated to 40 MeV, strikes a metal
foil.  The CTR is detected using a Golay detector attached to one arm
of a Michelson interferometer.  The alpha-magnet current and gun rf
phase are adjusted so as to maximize the CTR signal at the Golay
detector, which corresponds to the minimum bunch length.  The
interferometer is used to measure the autocorrelation of the CTR.  The
minimum phase approximation~\cite{Ref2} is used to derive the bunch
profile from the autocorrelation.  The high-brightness beam is
accelerated to 217 MeV and used to produce self-amplified spontaneous
emission (SASE) in five APS undulators installed in the Low- Energy
Undulator Test Line (LEUTL) experiment hall~\cite{Ref3}.  Initial
optical measurements showed a gain length of 1.3 m at 530 nm.
\end{abstract}

\section{INTRODUCTION}
\begin{figure}[htb]
\centering
\includegraphics*[width=80mm]{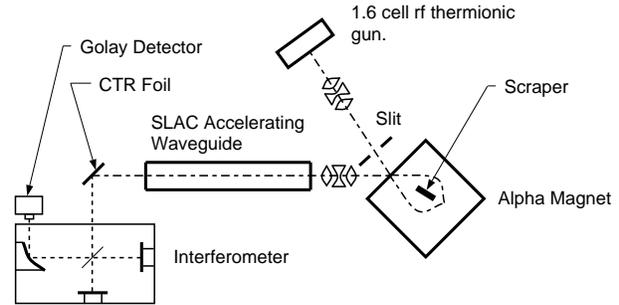}
\vspace{-1mm}
\caption{Layout of rf gun, alpha magnet, CTR apparatus, and linac beamline components.}
\label{fig1}
\end{figure}

The APS rf thermionic gun serves both as an injector for the APS
\cite{Ref1} storage ring as well as a high-brightness source for SASE
FEL experiments as part of the APS LEUTL project.  Tuning of the gun
as a high-brightness source was accomplished using CTR from the rf gun
beam accelerated to 40 MeV.  The gun, linac, and CTR setup are shown in
Figure~\ref{fig1}.  The beam emerges from the 1.6 cell $\pi$ mode rf
gun and proceeds to the alpha magnet via a beamline containing
focusing, steering, a kicker, and an entrance slit.  The alpha-magnet
vacuum chamber contains a scraper that is used to remove the low
energy/high emittance tail from the beam.  After the alpha-magnet, the
beam traverses some focussing and correction elements, then proceeds
through a 3-m SLAC s-band accelerating waveguide to the CTR foil.
The CTR foil is mounted on an actuator along with a YAG crystal, which
is used to focus the beam to a small beam spot at the foil position.
The CTR is collected by a lens and sent to a Michelson interferometer
with a Golay detector mounted on one arm.  The autocorrelation of the
CTR is performed by moving one arm of the interferometer while
recording the Golay detector output.  The Golay detector output can be
maximized at the peak of the autocorrelation scan and used to adjust
rf gun power and phase, beam current, and alpha-magnet current so as
to minimize the bunch length out of the alpha magnet.  Once this is
done, the CTR signal is a good relative measure of the bunch length.
\section{BEAM OPTIMIZATION}
\begin{figure}[htb]
\centering
\includegraphics*[width=85mm]{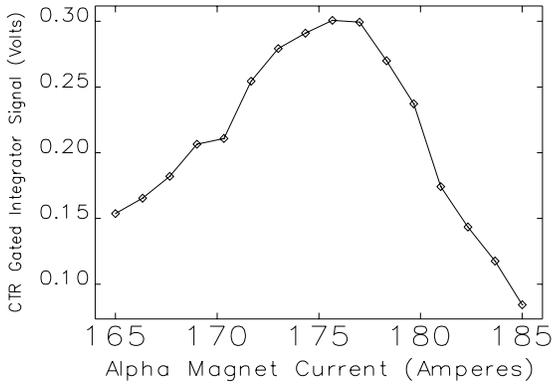}
\vspace{-5mm}
\caption{CTR gated integrator signal vs alpha magnet current.}
\label{fig3}
\end{figure}

To prepare the rf gun to produce a high-brightness beam one must first
scan the alpha-magnet current to find the minimum bunch length.
Typically the rf gun is powered anywhere from 1.5 to 1.7 MW, and the
heater current is adjusted to produce 1 to 2 nC in a train of 23
s-band bunches.  The gun power and beam current are kept constant
during the scans.  Prior to the scan, the beam is focused on the YAG
using quads before and after the alpha magnet, with the alpha magnet
``close'' to the setting required for minimum bunch length.  During
the scan, the rf gun phase must be adjusted linearly to compensate for
path length changes in the alpha magnet.  To maximize scan resolution,
the interferometer is set to maximize the Golay detector signal.
Figure~\ref{fig3} shows a typical alpha-magnet scan showing a peak at
175 amperes.  The curve represents the output of the Golay detector
from a gated integrator amplifier.

\begin{figure}[htb]
\centering
\includegraphics*[width=85mm]{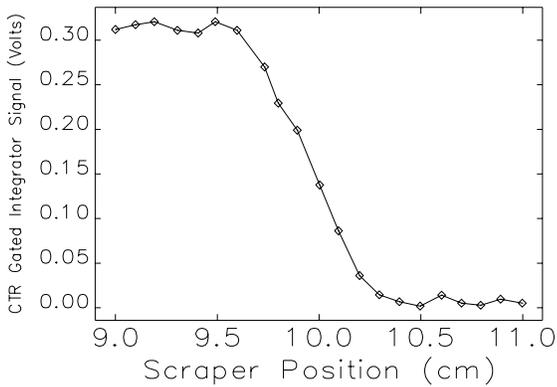}
\vspace{-5mm}
\caption{CTR gated integrator signal vs alpha-magnet scraper position.}
\label{fig4}
\end{figure}

\begin{figure}[htb]
\centering
\includegraphics*[width=85mm]{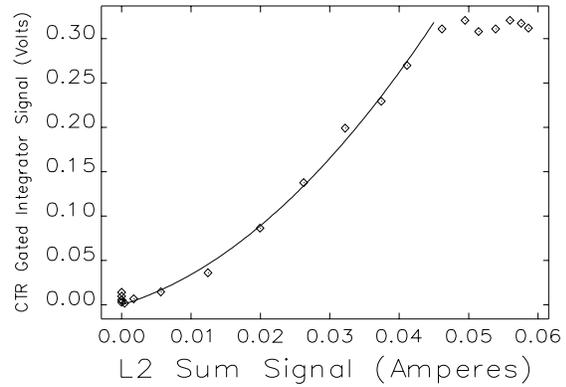}
\vspace{-5mm}
\caption{Plot of CTR gated integrator signal vs beam current as measured by a 
BPM.  The plot shows a quadratic fit along with the data indicating a strong 
quadratic dependence of the CTR.}
\label{fig5}
\end{figure}

Once the minimum bunch length has been found, an alpha-magnet scraper
scan is performed.  Simulations show a microbunch profile that has a
low-emittance, high-energy core beam and a high-emittance, low-energy
tail.  The scraper scan is performed to optimize removal of the
low-energy tail.  Figure~\ref{fig4} shows a typical scraper scan where
the CTR signal is plotted vs scraper position.  The edge of the core
beam is at approximately 9.5 cm.  Figure~\ref{fig5} shows a plot of
CTR signal vs beam current, as measured by a beam position monitor
(BPM) adjacent to the CTR foil, taken during the scraper scan.
Included with the data is a quadratic fit, showing the expected
quadratic dependence of the coherent radiation on the number of
particles.

\section{Bunch Profile Measurement}
\begin{figure}[htb]
\centering
\includegraphics*[width=85mm]{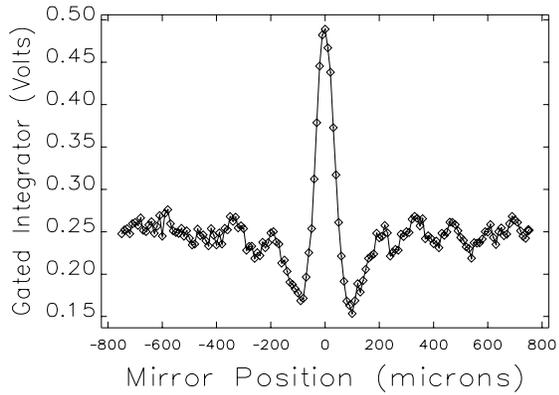}
\vspace{-5mm}
\caption{Autocorrelation of the gated integrator CTR signal.}
\label{fig6}
\end{figure}

\begin{figure}[htb]
\centering
\includegraphics*[width=85mm]{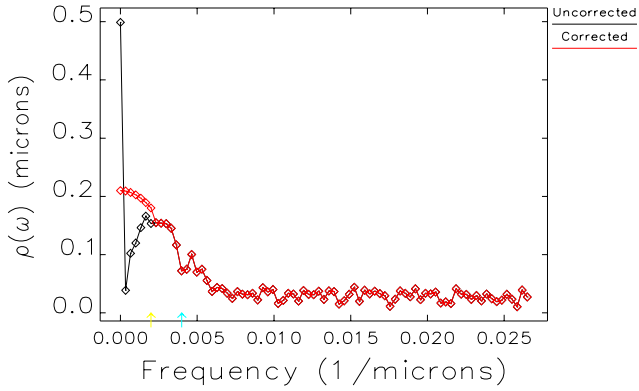}
\vspace{-5mm}
\caption{Amplitude spectrum derived from the autocorrelation and corrected spectrum at long wavelengths.}
\label{fig7}
\end{figure}

\begin{figure}[htb]
\centering
\includegraphics*[width=85mm]{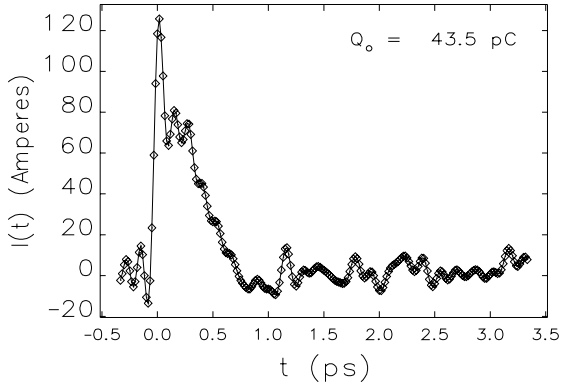}
\vspace{-5mm}
\caption{Bunch profile derived from corrected autocorrelation amplitude spectrum.}
\label{fig8}
\end{figure}

Once the scraper position is determined, the interferometer is used to
measure the autocorrelation of the digitized gated integrator CTR
signal.  Figure~\ref{fig6} shows the autocorrelation measured for a
beam of 1 nC in 23 S-band micropulses.  Autocorrelation processing
begins with taking the fast Fourier transform (FFT) of the
autocorrelation, which gives the square of the bunch spectrum.  The
method of Lai and Sievers is then used to reconstruct the phase
spectrum from the amplitude spectrum by computing a principal value
integral.  Once the phase spectrum is obtained, an inverse FFT is
performed to derive the microbunch profile.
Additional processing is performed to correct for the reduced response
of the Golay detector at low frequencies (long wavelengths).  Since
any bunch spectrum approaches low frequencies quadratically, a
quadratic fit is performed for frequencies from the Golay detector 3-dB 
point to a user-selectable higher frequency, typically including 3
to 5 frequency points~\cite{Ref2}.  The fit is then used to
extrapolate quadratically to DC from the Golay detector 3-dB point.
Figure~\ref{fig7} shows the amplitude spectrum derived from the measured
autocorrelation and the corrected spectrum for low frequencies.
The main effects of this low-frequency correction is to broaden the
derived bunch profile and flatten the dips in the autocorrelation
adjacent to the peak.  These dips are unphysical since the
autocorrelation is always positive.  Figure~\ref{fig8} shows the derived
bunch profile from the the corrected autocorrelation spectrum.  The
overall profile contains a high-current peak ($>$ 100 amperes), a
lower current shoulder, and is overall about 400 fs wide.
This beam was used for SASE measurements.

\section{Measurement of SASE Gain}

\begin{figure}[htb]
\centering
\includegraphics*[width=85mm]{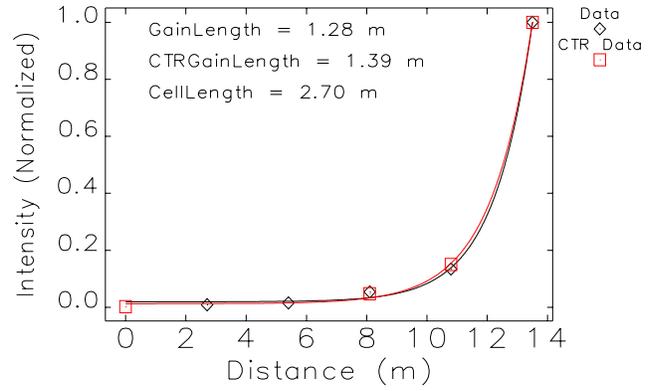}
\vspace{-5mm}
\caption{SASE gain measured at undulator diagnostics stations.}
\label{fig9}
\end{figure}

The beam prepared as described above was accelerated to 217 MeV.  The
emittance was measured in the transport line using the standard
three-screen technique, giving a normalized emittance of approximately
10 $\pi$ mm.  The energy spread is estimated to be ~0.1\%.  The beam was
transported to the undulator hall and passed through five APS
undulators with diagnostics stations between them.  Figure~\ref{fig9}
shows the measured photon intensity (corrected for spontaneous
background) at each undulator diagnostic station.  The solid line is
an exponential fit to the data showing a gain length of 1.3 m for both
undulator radiation and coherent transition radiation
data~\cite{Ref4}, in agreement with a calculation using the previously
listed peak current, emittance, and energy spread.

The rf thermionic gun beam was quite stable once tuning was completed.
One limitation of the beam is that the microbunch length is on the
order of the electron slippage length.  The final saturated power is
therefore expected to be lower for this beam.

\section{Acknowledgements}
The authors thank J. Lewellen, S. Milton, and J. Galayda for useful
comments and suggestions.

\end{document}